\documentstyle[sprocl,epsf]{article}

\begin{document}

\title{EVIDENCE FOR THE OBSERVATION OF A GLUEBALL}

\author{DON WEINGARTEN}

\address{IBM Research, Yorktown Heights\\
NY 10598, USA}

\maketitle\abstracts{
I briefly review lattice QCD, the valence (quenched) approximation, and
the application of both to the determination of the mass and two-body
decay couplings of the lightest scalar glueball.  Results in agreement
with the observed properties of $f_J(1710)$ strongly suggest this resonance
is largely a scalar glueball.}

\section{Introduction}\label{sect:intro}

The existence of chromoelectric field is the key hypothesis QCD adds to
the quark model. The observation in experiment of a glueball would be a
direct confirmation of the existence of chromoelectric field. So it
would be nice to find one. Identifying a glueball in experiment is
tough, however.  The kinematics of QCD provides no clear footprint for
glueballs comparable, say, to the footprint the Weinberg-Salam model
gives for the W's and the Z.  The lightest states, which would be
easiest to detect, are not expected to differ drastically in mass or
decay properties from mesons containing quarks and antiquarks.  On the
other hand, $J^{PC}$ combinations which are impossible for
quark-antiquark states but which can occur for heavier glueball
excitations can also be realized with multiple quark-antiquark pairs.  A
reliable calculation of the consequences of QCD's dynamics for the
masses and decay properties of glueballs appears to be a necessary
element in the identification of glueballs in experiment. At present
such calculations can be done only numerically, using the lattice
formulation of QCD in combination with the valence approximation.

As of three years ago, two independent calculations had been completed
of the infinite-volume, continuum limit of the mass of the lightest
glueball, which turns out to be a scalar. Hong Chen, Jim Sexton,
Alessandro Vaccarino and I obtained \cite{Vaccarino} a value of 1740(71)
MeV using ensembles of 25000 to 30000 gauge configurations on each of
several different lattices.  An earlier valence approximation
calculation by the UKQCD-Wuppertal collaboration~\cite{Livertal}, when
extrapolated to the continuum limit \cite{Weingarten94}, yields 1625(94)
MeV for the lightest scalar glueball mass.  This calculation used
several different lattices with ensembles of between 1000 and 3000
configurations each.  If the two mass evaluations are combined, taking
into account the correlations between their statistical uncertainties
arising from a common procedure for converting lattice quantities into
physical units, the result is 1707(64) MeV for the scalar glueball mass.
Both the mass prediction with larger statistical weight and the combined
mass prediction agree with the mass of $f_J(1710)$ and are inconsistent
with all but $f_0(1500)$ among the established flavor
singlet resonances which could be scalars.  For $f_0(1500)$ the
disagreement is still by more than three standard deviations.  In
addition, observed mass values for other scalar quark-antiquark states
suggest a mass near 1500 MeV for the lightest $s\overline{s}$ scalar.

These calculations by themselves, however, do not make a strong case for
taking $f_J(1710)$ as a scalar glueball.  The key question which the
mass results do not answer is whether the lightest scalar glueball has a
decay width small enough for this particle actually to be identified in
experiment.  If the scalar glueball had a width of a GeV or more, the
prospect of ever finding one seems remote.  But a scalar glueball with a
width of a few hundred MeV or less and mass in the neighborhood of 1700
MeV would be hard to miss and should already have been seen in
experiment.  A further question in the identification of $f_J(1710)$ as
a glueball is raised by the argument that since glueballs are flavor
singlets they should have the same decay couplings to $\pi_0 \pi_0 $, to
$K_S K_S$, and to $\eta \eta$. This equality is violated by $f_J(1710)$
decay couplings.

To address these questions, Jim Sexton, Alessandro Vaccarino and
I~\cite{glbdecay} have calculated the decay coupling constants of the
lightest scalar glueball to pairs of pseudoscalar mesons. The
calculation was done in the valence approximation on a lattice $16^3
\times 24$ with $\beta$, defined as $6/g^2$, set at 5.70, 
corresponding to lattice spacing, determined from the $\rho$ mass, of
about 0.15 fm and lattice period of about 2.3 fm. For the total width of
the scalar glueball to pairs of pseudoscalar quark-antiquark states we
obtained $108 \pm 29$ MeV.  The combined correction for the errors in our
prediction arising from the valence approximation, from finite lattice
spacing and from finite lattice volume we believe would be 
less than 50\%. Based on our value for
two-body decays, any reasonable guess for the partial width for multibody
decay modes leads to a total width small enough for the scalar glueball to
be hard to miss in experiment.  In fact, the observed~\cite{Lind} width
of $f_J(1710)$ into pairs of pseudoscalars is $99 \pm 15$ MeV,
consistent with our result. We obtain also a violation of the expected
equality of glueball decay rates to $\pi_0 \pi_0 $, to $K_S K_S$, and to
$\eta \eta$ in agreement with the observed results for $f_J(1710)$.

Although so far I have simplified the story by supposing that physical
resonances are either entirely glueballs or entirely quark-antiquark,
another possibility is a state which is a linear combination of a
glueball and a quark-antiquark state.  In the valence approximation,
however, glueballs contain no admixture of configurations with valence
quarks or antiquarks.  Thus we consider the agreement between the mass
and decay couplings found in the valence approximation and the observed
mass and decay couplings of $f_J(1710)$ to be strong evidence that this
state is largely a scalar glueball with at most some relatively smaller
amplitude for configurations including valence quark-antiquark pairs.

The glueball calculations presented here were carried out on the GF11
parallel computer \cite{Weingarten90} at IBM Research and took
approximately 30 months to complete at a sustained computation rate of
between 6 and 7 Gflops.

In the remainder of this talk, I will briefly review lattice QCD, the
valence approximation, the glueball mass calculation, the decay
calculation, and close with a more detailed comparison of the lattice
predictions with experiment.

\section{Lattice QCD}

Lattice QCD, as its name implies, approximates continuous space-time (at
negative imaginary values of time) by a discrete lattice including only
a finite number of points.  Predictions for the real world of continuous,
infinite-volume space-time are supposed to be recovered from lattice QCD
by taking the limit of lattice predictions as lattice spacing goes to
zero and lattice volume to infinity.  It is convenient, though not
necessary, to take the lattice to be hypercubic with periodic boundary
conditions.  Living at each site are lattice versions of the
chromoelectric potential, $A^j_{\mu}(x)$, given by real numbers, and the
quark and antiquark fields, $\overline{\Psi}_{sd}(x)$ and
$\Psi_{sc}(x)$, respectively, given by Grassmann variables.  Here $x$ is
a lattice site, $j$, $c$ and $d$ are color indices for the adjoint,
fundamental and conjugate representations, respectively, $\mu$ is a
lattice direction and $s$ is a spin index.

Vacuum expectation values of time ordered products of fields are then
defined by a path integral. A typical vacuum expectation value, which I
offer in place of the slightly more complicated general rule, is
\begin{eqnarray}
\label{vacexamp}
< \overline{\Psi}(x) \Psi(x) >  & = &   
 Z^{-1} \int \prod d\mu_{A} \int \prod d \Psi d \overline{\Psi} \: 
 \overline{\Psi}(x) \Psi(x) exp( {\cal S}_A + {\cal S}_{\Psi}), \nonumber \\ 
Z & = & 
  \int \prod d\mu_{A} \int  \prod d \Psi d \overline{\Psi} 
exp( {\cal S}_A + {\cal S}_{\Psi}), \\ 
{\cal S}_A & = & -\frac{1}{4 g^2} \int d^4 x F_{\mu \nu}^j F_{\mu
\nu}^j. \nonumber \\
{\cal S}_{\Psi} & = & 
\int d^4 x \overline{\Psi} (p\!\!\!/ - A\!\!\!/ + i m) \Psi, \nonumber
\end{eqnarray}
where $\int \prod d\mu_{A}$ can be thought of, but is not quite, a
product over all $x$, $\mu$ and $j$ of the integral
$\int_{-\infty}^\infty d A^j_{\mu}(x)$ and $\int \prod d \Psi d
\overline{\Psi}$ is a product of Grassmann integration on each
$\overline{\Psi}_{sd}(x)$ and $\Psi_{sc}(x)$.  For the continuum
quantities and integrals which appear in Eqs.~(\ref{vacexamp}), I
actually intend lattice approximations but use the continuum expressions
as more recognizable alias's.

With respect to the quark and antiquark fields, Eqs.~(\ref{vacexamp})
are a Grassmann version of gaussian integrals. The integrals can be done
analytically and give
\begin{eqnarray}
\label{matsalam}
< \overline{\Psi}(x) \Psi(x) >  & = & Z^{-1} \int \prod d\mu_{A} \: 
 tr[ (p\!\!\!/ - A\!\!\!/ + i m)^{-1}( x, x)] \times \nonumber \\
& & det(p\!\!\!/ - A\!\!\!/ + i m) \,  exp( {\cal S}_A),  \\ 
Z  & = &   \int \prod d\mu_{A}  \: 
 det(p\!\!\!/ - A\!\!\!/ + i m) \, exp( {\cal S}_A), \nonumber
\end{eqnarray}
where the trace is with respect to spin and color indices.

The lattice QCD industry does integrals like those in
Eq.~(\ref{matsalam}) by Monte Carlo.
With a large enough ensemble of random $A$ fields, 
$[A_{\mu}^j(x)]_1, \ldots [A_{\mu}^j(x)]_N$, generated by computer
according to the differential probability
\begin{equation}
\label{dnu}
d \nu = Z^{-1}\prod d\mu_{A} \: 
det(p\!\!\!/ - A\!\!\!/ + i m) \,  exp( {\cal S}_A).  
\end{equation}
the vacuum expectation value
of Eqs.~(\ref{matsalam}) becomes
\begin{equation}
\label{mcvac}
< \overline{\Psi}(x) \Psi(x) >  = 
\frac{1}{N} \sum_k tr[ (p\!\!\!/ - A\!\!\!/_k + i m)^{-1}( x, x)]
\end{equation}

\section{The Valence Approximation}

The problem in generating, with present computer hardware, ensembles of
$A$ according to $d \nu$ of Eq.~(\ref{dnu}), and the origin of the
valence approximation, is the factor $det(p\!\!\!/ - A\!\!\!/ + i m)$.
For the moment forget about this factor.  A simple algorithm to generate
an ensemble of $A$ begins with some arbitrarily chosen field, for
example all $A$ components set to 0, and then successively walks across
all sites $x$ and directions $\mu$ on the lattice modifying the vector
of $A^j_{\mu}(x)$, for all $j$, at each. The collection of $A$ resulting
from each complete sweep updating all $x$ and $\mu$ gives a possible
Monte Carlo ensemble. The update at each $x$ and $\mu$ proceeds in two
steps.  First a random new trial vector of $A^j_{\mu}(x)'$, for all $j$,
is generated according to a rule the details of which I will skip. Then
the vector of $A^j_{\mu}(x)'$ is either installed in place of the old
$A^j_{\mu}(x)$ or thrown away with a probability depending on the change
$\Delta {\cal S}_A$ which this replacement would cause in ${\cal
S}_A$. Now, the action ${\cal S}_A$ approximates the derivatives in
$F_{\mu \nu}^j(x)$ with nearest-neighbor differences. So computing the
change $\Delta {\cal S}_A$ arising from the trial replacement of
$A^j_{\mu}(x)$ by $A^j_{\mu}(x)'$ leads to work involving only
$A^k_{\nu}(y)$ for $y$ equal to $x$ or nearest-neighbors of $x$.  Only a
fixed number of arithmetic operations are required independent of the
size of the lattice.  The total work to update $A^j_{\mu}(x)$ for all
$x$ and $\mu$ to produce a new member of the Monte Carlo
ensemble is therefore proportional to $V$, the number of sites in the
lattices.

Now put $det(p\!\!\!/ - A\!\!\!/ + i m)$ back in $d \nu$ in
Eq.~(\ref{dnu}).  The algorithm for generating a Monte Carlo ensemble is
nearly unchanged, except that in place of the the change $\Delta {\cal
S}_A$, we need the change $\Delta {\cal S}_A +
\Delta \log[ det(p\!\!\!/ - A\!\!\!/ + i m)]$. Since $det(p\!\!\!/ -
A\!\!\!/ + i m)$ couples together all $A^j_{\mu}(x)$ on the entire
lattice, the work to find $\Delta {\cal S}_A +
\Delta \log[ det(p\!\!\!/ - A\!\!\!/ + i m)]$ in the update for a single
$x$ and $\mu$ is already of order $V$. The work to update all sites
becomes of order $V^2$. But recall, for a moment, that predictions for
the real world are found from lattice QCD predictions by taking limits
of zero lattice spacing and infinite lattice volume. Even a rough
approximation to these limits, at least for the simplest lattice
implementations of ${\cal S}_A$ and ${\cal S}_{\Psi}$ in
Eqs.~(\ref{vacexamp}), requires lattice dimensions of at least $10 \times
10 \times 10 \times 10$ thus $V$ of $10^4$.  The cost of including
$det(p\!\!\!/ - A\!\!\!/ + i m)$ therefore becomes a factor of $10^4$ or
more. Actually, it turns out that somewhat fancier algorithms than what
I just described handle $det(p\!\!\!/ - A\!\!\!/ + i m)$ at lower cost,
perhaps a factor of 100 to 1000 for the largest lattices. Unfortunately,
present computer power is just barely sufficient to handle lattices
large enough to give infinite-volume, continuum limit predictions in the
absence of $det(p\!\!\!/ - A\!\!\!/ + i m)$ in $d \nu$. The increase by
a factor of 100 to 1000 in work required to include $det(p\!\!\!/ -
A\!\!\!/ + i m)$ in $d \nu$ is fatal.

What to do?  A possible answer~\cite{Weingarten82} is suggested by
viewing $det(p\!\!\!/ - A\!\!\!/ + i m)$ through the eyes of weak
coupling perturbation theory.  In weak coupling perturbation expansions,
$det(p\!\!\!/ - A\!\!\!/ + i m)$ gives rise to closed quark loops
interrupting gluon lines inside diagrams. So $det(p\!\!\!/ - A\!\!\!/ +
i m)$ is in charge of the QCD analogue of the particle-hole polarization
process that occurs when an electromagnetic field propagates through a
solid. In the case of electrodynamics, we know that for sufficiently
weak, sufficiently low momentum electromagnetic fields the effect of
particle-hole polarization is accurately approximated by simply
replacing all charges and fields by screened values obtained by dividing
each by the solid's dielectric constant. A plausible hypothesis is that
the same holds for QCD. Namely, for processes involving sufficiently
weak, sufficiently low momentum chromoelectric fields, we can omit
$det(p\!\!\!/ - A\!\!\!/ + i m)$ from $d \nu$ and just replace the
chromoelectric charge $g$ in ${\cal S}_A$ of Eqs.~(\ref{vacexamp}) with a
screened charge $g/\eta$, for some QCD analogue dielectric constant
$\eta$. This replacement is the valence approximation, also called the
quenched approximation.

\begin{figure}
\begin{center}
\leavevmode
\epsfxsize=65mm
\epsfbox{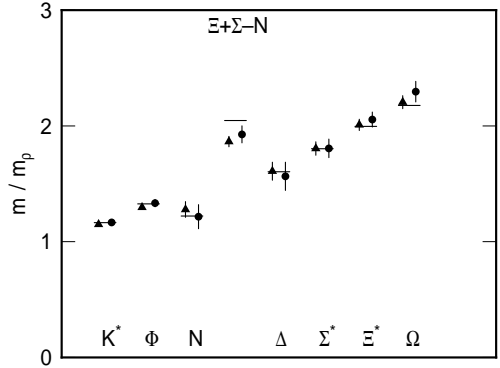}
\vskip -2mm
\end{center}
\caption{ 
Continuum limit valence approximation hadron mass predictions
in finite volume, triangles, and extrapolated to infinite volume, circles.}
\label{fig:mass2}
\vskip -2mm
\end{figure}

In part as a test of the method, a calculation~\cite{Butler} has been
done of the infinite-volume, continuum limit of the valence
approximation to light hadron masses. A total of 11 masses were
calculated, three of which were used to determine input parameters. The
$\pi$ mass was used to set the up and down quark masses, taken to be
equal. The K mass was used to set the strange quark mass. The $\rho$
mass was used to determine the gauge coupling constant. QCD's supply of
free parameters is then exhausted.  The remaining eight masses are
predictions. The results in units of the $\rho$ mass are shown in
Figure~\ref{fig:mass2}. The triangles give zero-lattice-spacing results
in a box with period of about 2.3 fm, which turns out to be nearly
infinite volume for light hadron masses. The circles show infinite-volume
predictions obtained by applying an additional correction to the
2.3 fm results. Out of eight numbers, the biggest disagreement between
prediction and experiment is 6\%. The statistical uncertainties in these
numbers range up to 8\%, however. For eight numbers with these
uncertainties, one disagreement with experiment by 6\% is expected.
Thus 6\% should be treated as the one sigma upper bound on the error in
the valence approximation.

So taking the valence approximation as accurate to within 6\% for light
hadron masses, we applied it to predicting glueball properties.

\section{Scalar Glueball Mass}

To determine the scalar glueball mass~\cite{Vaccarino} we evaluated the
vacuum expectation value
\begin{eqnarray}
\label{glueprop}
C( t) & = & < g(t) g(0) >, \nonumber \\
g(t) & = & h(t) - < h(t)>, \\
h(t) & = & \int d^3 x \: F_{ab}^j( \vec{x}, t) \, F_{ab}^j( \vec{x},t),
\nonumber
\end{eqnarray}
where $a$ and $b$ are summed only over space directions, and as
before, I actually intend lattice approximations for $F_{ab}^j( \vec{x},
t)$ and $\int d^3 x$ but use the continuum expressions as more
recognizable alias's. By inserting a complete set of energy eigenstates
between the two glueball operators in $C( t)$, it is easy to show that
for large values of the euclidean time variable
$t$, $C( t)$ has the asymptotic behavior
\begin{equation}
\label{glueasym}
C( t)  \rightarrow 
< vac | g( 0) | g> < g | g( 0) | vac > exp( -m_g t) + \ldots ,
\end{equation}
where $| g>$ is the zero-momentum state of the lightest scalar glueball,
$m_g$ is its mass, and the omitted terms come from scalar glueball
excitations and fall off exponentially in $t$ with a coefficient larger than
$m_g$. So for large $t$, the effective glueball mass $m_g( t)$ has
asymptotic behavior
\begin{equation}
\label{effmass}
m_g( t) = \log \frac{C( t)}{C( t + 1)} \rightarrow m_g.
\end{equation}

\begin{figure}
\begin{center}
\leavevmode
\epsfxsize=70mm
\epsfbox{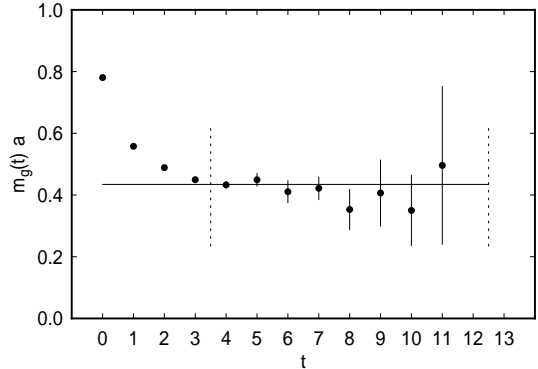}
\vskip -2mm
\end{center}
\caption{ 
Scalar glueball effective mass on a lattice $30 \times 32^2
\times 40$ with lattice spacing of 0.050 fm.}
\label{fig:x32s}
\vskip -2mm
\end{figure}

Figure~\ref{fig:x32s} shows $m_g( t)$ in lattice units, for a lattice
$30 \times 32^2 \times 40$ at $\beta$ of 6.4, obtained from an ensemble
of 25440 configurations of chromoelectric field.  The lattice spacing in
this case, determined from the $\rho$ mass, is about 0.05 fm and lattice
period is about 1.6 fm. For $t$ of 4 and greater, $m_g( t)$ is
consistent with a constant shown by the horizontal line, giving
according to Eq.~(\ref{effmass}) a value for $m_g$.  From similar
calculations for a range of different lattice periods, we found that 1.6
fm gives results within a fraction of a percent of the infinite-volume
limit.

\begin{figure}
\begin{center}
\leavevmode
\epsfxsize=70mm
\epsfbox{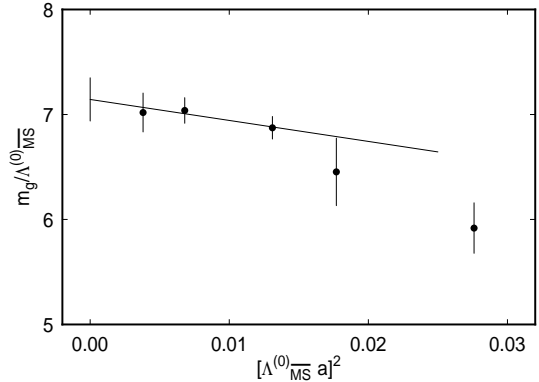}
\vskip -2mm
\end{center}
\caption{ 
Continuum limit of the scalar glueball mass.}
\label{fig:a0}
\vskip -2mm
\end{figure}

To obtain the zero-lattice-spacing limit of $m_g$, we evaluated $C( t)$
for five different values the lattice spacing, in all cases with lattice
period of 1.6 fm or greater.
Now, the lattice version of the action
$S_A$, defined in Eqs.~(\ref{vacexamp}), which occurs in a valence 
approximation calculation of $C( t)$,
replaces derivates by symmetric finite difference, obeying
\begin{equation}
\label{finitediff}
\frac{f( x + a) - f( x - a)}{2 a} = 
\frac{d f(x) }{d x} +
\frac{a^2}{6} \frac{d^3 f(x)}{d x^3} + \ldots,
\end{equation}
where $a$ represents lattice spacing. The error in replacing an integral
by a sum in $S_A$ goes to zero more rapidly than $a^2$.  Thus for small
$a$, $S_A$ will be the continuum action plus an error of order $a^2$. To
leading order in $a$, $m_g$ measured in physical units will then also be
its continuum value plus an error of order $a^2$.  Figure~\ref{fig:a0}
shows the five values of $m_g$ in physical units as a function of $a^2$
in physical units. The unit $\Lambda^{(0)}_{\overline{MS}}$ in
Figure~\ref{fig:a0} is the valence approximation
$\Lambda_{\overline{MS}}$, for which a value of $243.7
\pm 6.8$ MeV was obtained in the course of our calculation of
quark-antiquark meson masses~\cite{Butler}.  To convert $m_g a$ and $a$
to physical units, values of $\Lambda^{(0)}_{\overline{MS}} a$ in
lattice units were found by the two-loop Callan-Symanzik equation.

As it turns out, the $\rho$ mass in lattice units $m_{\rho} a$
scales almost perfectly with $\Lambda^{(0)}_{\overline{MS}} a$ for the
range of parameters in Figure~\ref{fig:a0}.  So Figure~\ref{fig:a0} can
also be thought of simply as a plot of $m_g$ and $a$ in units of
$m_{\rho}$ but with the axes mislabeled by powers of $243 MeV /
m_{\rho}$.
 
The three points at smallest lattice spacing in Figure~\ref{fig:a0} fit
a straight line in $a^2$ quite well. Extrapolating to zero lattice 
spacing gives the limiting $m_g$ of 1740(71) MeV, in good agreement with
the observed mass of $f_J(1710)$, as mentioned earlier in
Section~\ref{sect:intro}. This result is rather insensitive to 
how the extrapolation is done. Even if we had arbitrarily taken
either of the last two points as the continuum limit, a procedure which
is certainly less reliable than extrapolation, the answer would change
by less than half of the 71 MeV statistical uncertainty.

\section{Decay Couplings}

We calculated~\cite{glbdecay} coupling constants both for the decay of
the lightest scalar glueball to pairs of pseudoscalars at rest and to
pairs of pseudoscalars with oppositely directed momenta of magnitude $2
\pi/ L$, for lattice period $L$. For the sake of simplicity, I am only
going to give details for the decay to pseudoscalars at rest.  To
measure the decay at 
rest we evaluated the three-point function
\begin{eqnarray}
\label{decayprop}
C(t_g, t_{\pi}) & = & < g( t_g) \sum_f \pi^{\dagger}_f( t_{\pi}) \pi_f( 0)>, \\
\pi_f( t) & = & \int d^3 x \overline{\Psi}_u( \vec{x}, t) \gamma_5
\lambda_f \Psi_d(
\vec{x}, t), \nonumber
\end{eqnarray}
where $g( t)$ is defined in Eqs.~(\ref{glueprop}) and the $\lambda_f$ are
an orthonormal set of SU(3) flavor matrices.  The $u$, $d$ and $s$ quark
masses were set equal, and chosen so that the energy of the lightest
flavor-singlet, two-pseudoscalar, zero-momentum state coincides
with the glueball mass $m_g$.  How to apply this calculation to the
real world with a very different set of quark masses, I will explain in the
next section.

By inserting a complete set of energy eigenstates
between $g( t_g)$ and $\pi^{\dagger}_f( t_{\pi})$ in $C(t_g, t_{\pi})$,
you can show that at large $t_g - t_{\pi}$
$C( t_g, t_{\pi})$ has the asymptotic behavior
\begin{eqnarray}
\label{decayasym}
\lefteqn{C( t_g, t_{\pi}) \rightarrow } \\
& & < vac | g( 0) | \pi \pi> 
< \pi \pi | \sum_f \pi^{\dagger}_f( t_{\pi}) \pi_f( 0)  | vac > exp(
-m_g t_g) 
+ \ldots , \nonumber
\end{eqnarray}
where $| \pi \pi>$ is the lightest zero-momentum, flavor-singlest
state of two pseudoscalars and the terms omitted fall off
exponentially in $t_g$ with coefficients larger than $m_g$.

The coupling constant we are looking for can be found
from $< vac | g( 0) | \pi \pi>$.
To obtain $< vac | g( 0) | \pi \pi>$ from $C(t_g, t_{\pi})$ using
Eq.~(\ref{decayasym}), we need the value of $< \pi \pi | \sum_f
\pi^{\dagger}_f( t_{\pi}) \pi_f( 0) | vac >$.
This factor carries the corrections to the decay calculation
arising from $\pi-\pi$ final state interactions. 
We extracted $< \pi \pi | \sum_f
\pi^{\dagger}_f( t_{\pi}) \pi_f( 0) | vac >$
from the four-point function
\begin{equation}
\label{fourpoint}
C( t_{\pi 3}, t_{\pi 2}, t_{\pi 1}) = 
<  \sum_f \pi^{\dagger}_f( t_{\pi}) \pi_f( 0) 
\sum_g \pi^{\dagger}_g( t_{\pi}) \pi_g( 0) >.
\end{equation}
The job of evaluating $C( t_{\pi 3}, t_{\pi 2}t_{\pi 1})$ and extracting
$< \pi \pi | \sum_f
\pi^{\dagger}_f( t_{\pi}) \pi_f( 0) | vac >$ I will not describe here.
If we had simply taken for $< \pi \pi | \sum_f
\pi^{\dagger}_f( t_{\pi}) \pi_f( 0) | vac >$ its value assuming no
$\pi-\pi$ interaction, however, the final coupling constant would be
changed by less than 15\%.

As a byproduct of finding the four-point function $C( t_{\pi 3},
t_{\pi 2}t_{\pi 1})$, we were also able to determine the
first omitted term in Eq.~(\ref{decayasym}).  To accelerate the approach
to large $t_g - t_{\pi}$ asymptopia, we subtracted this term from $C(
t_g, t_{\pi})$, giving $D( t_g, t_{\pi})$.  
The large $t_g - t_{\pi}$
asymptotic behavior of $D( t_g, t_{\pi})$ is then
\begin{equation}
\label{decayasym1}
D( t_g, t_{\pi}) \rightarrow 
\lambda K( t_g, t_{\pi}),
\end{equation}
where $\lambda$ is the decay coupling constant for $g \rightarrow \pi
\pi$ at rest and $K( t_g, t_{\pi})$ is a kinematic factor determined
in part from $C( t_{\pi 3}, t_{\pi 2}, t_{\pi 1})$.
Alternatively,
defining from $D( t_g, t_{\pi})$ an effective $\lambda( t_g, t_{\pi})$,
\begin{equation}
\label{efflam}
\lambda( t_g, t_{\pi}) = \frac{D( t_g, t_{\pi})}{ K( t_g, t_{\pi})},
\end{equation}
it follows that for large $t_g - t_{\pi}$, $\lambda( t_g, t_{\pi})$
approaches the decay coupling constant $\lambda$.  

\begin{figure}
\begin{center}
\leavevmode
\epsfxsize=65mm
\epsfbox{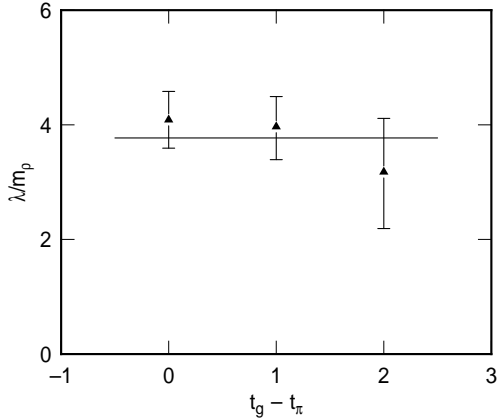}
\vskip -2mm
\end{center}
\caption{ 
$\lambda$ as function of $t_g - t_{\pi}$.}
\label{fig:g0}
\vskip -2mm
\end{figure}

Values of $\lambda( t_g, t_{\pi})$ were calculated on a lattice $16^3
\times 24$ at $\beta$ of 5.70, 
corresponding to lattice spacing about 0.15 fm and lattice period of
about 2.3 fm, from an ensemble of 10500 configurations of $A$ field.
For $t_g - t_{\pi}$ fixed, we found $\lambda( t_g, t_{\pi})$ independent
of $t_{\pi}$ for $t_{\pi} \ge 3$. Figure~\ref{fig:g0} shows $\lambda(
t_g, t_{\pi})$, for $t_{\pi} \ge 3$, as a function of $t_g - t_{\pi}$.
Largely as a consequence of the subtraction in defining $D( t_g,
t_{\pi})$, the data is consistent with a constant from $t_g - t_{\pi}$
of zero on out.  This constant is the final value of $\lambda$ for
glueball decay to pseudoscalars at rest.

\section{Comparison with Experiment}

\begin{figure}
\begin{center}
\leavevmode
\epsfxsize=65mm
\epsfbox{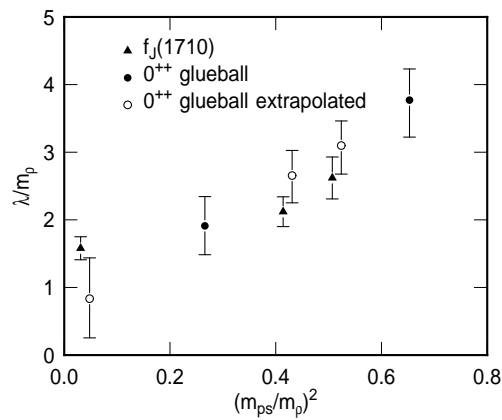}
\vskip -2mm
\end{center}
\caption{ 
Decay couplings.}
\label{fig:lambdas}
\vskip -2mm
\end{figure}

So far I have given u, d and s quarks degenerate, unphysical mass
values. Here is how to fix that.  An expansion to first order in the
quark mass matrix taken around some relatively heavy SU(3) symmetric
point gives glueball decay couplings to $\pi$'s, K's and the $\eta$'s
which are a common linear function of each meson's average quark
mass. Since meson masses squared are also nearly a linear function of
average quark mass, the decay couplings are a linear function of meson
masses squared.  Therefore, a linear fit to our predictions for decay
couplings as a function of pseudoscalar mass squared lets you
extrapolate from unphysical degenerate values of quark masses to
physical nondegenerate values of quark masses.  Figure~\ref{fig:lambdas}
shows predicted coupling constants as a function of predicted meson mass
squared along with a linear extrapolation of the predicted values to the
physical $\pi$, K and $\eta$ masses. Shown also are the observed
couplings\cite{Lind} for decays of $f_J(1710)$ to pairs of $\pi$'s, K's
and $\eta$'s.  Everything is in units of the $\rho$ mass.  The total
predicted width for glueball decay to pseudoscalar pairs becomes 108(28)
MeV, in comparison to 99(15) MeV for $f_J(1710)$.

How far are the valence approximation, finite lattice spacing decay
couplings likely to be from the real world? From the comparison of
finite lattice spacing valence approximation hadron masses with their
values in the real world, I would expect an error of 15\% or less in
going to the continuum limit and another 6\% or less arising from the
valence approximation.  The total predicted width for glueball decay to
two pseudoscalars should then have an error of less than 50\%. A 50\%
increase in our predicted two-body decay width, combined with any
reasonable corresponding guess for multibody decays, gives a total
glueball width small enough for the particle to be observed easily.

For the continuum limit glueball mass, a 6\% valence approximation error
would be 100 MeV, but according to an adaptation of an argument giving a
negative sign for the valence approximation error in
$f_{\pi}$~\cite{Butler}, the sign of this error is also expected to be
negative.  Thus the the only established resonance aside from
$f_J(1710)$ with the correct quantum numbers and mass close enough to
1740 to be a candidate~\cite{Amsler2} for the scalar glueball is
$f_0(1500)$.  The most likely interpretation of $f_0(1500)$, however, I
think is as an $s\overline{s}$ quark-antiquark meson.  The
$s\overline{u}$ scalar and tensor are nearly degenerate at about 1430
MeV. So the $s\overline{s}$ scalar and tensor should lie close to each
other somewhere above 1430 MeV. The $s\overline{s}$ tensor has been
identified at 1525 MeV.  An $s\overline{s}$ scalar around 1500 MeV seems
to me hard to avoid.

\end{document}